\documentclass[a4paper,11pt,epsfig,clip,here,twoside,ams]{article}

\usepackage{amsmath}
\usepackage{amssymb}
\usepackage{epsfig}

\newcommand{\interskip}{\bigskip}

\newcommand{\hide}[1]{}
\newcommand{\Fig}[1]{Fig.~\ref{fig:#1}}

\newcommand{\eq}[1]{eq.~(\ref{eq:#1})}
\newcommand{\Eqs}[1]{Eqs.~(\ref{eq:#1})}
\newcommand{\eqs}[1]{eqs.~(\ref{eq:#1})}
\newcommand{\fracwithdelims}[4]{\left#1 \frac{#3}{#4} \right#2}
\newcommand{\dm}[1]{{\Delta m^2_{#1}}}
\newcommand{\M}{M}
\newcommand{\Meff}{{M_{\text{eff}}}}
\newcommand{\V}{V}

\newcommand{\Eres}{E_{\text{res}}}
\newcommand{\ENP}{E_{\text{NP}}}

\newcommand{\GeV}{\,\mathrm{GeV}}

\newcommand{\eV}{\,\mathrm{eV}}
\newcommand{\cm}{\,\mathrm{cm}}

\newcommand{\ete}{\epsilon_{\tau e}}
\newcommand{\etu}{\epsilon_{\tau \mu}}
\newcommand{\eue}{\epsilon_{\mu e}}
\newcommand{\ve}{\varepsilon}

\newlength{\myem}
\newcommand{\sep}[1]{#1}
\newcounter{mysubequation}[equation]
\renewcommand{\themysubequation}{\alph{mysubequation}}
\newcommand{\mytag}{\stepcounter{mysubequation}%
\tag{\theequation\protect\sep{\themysubequation}}}
\newcommand{\globallabel}[1]{\refstepcounter{equation}\label{#1}}

\makeatletter
\renewcommand{\section}{\@startsection{section}{1}{0em}%
        {-3.5ex \@plus -1ex \@minus -.2ex}%
        {2.3ex \@plus.2ex}%
        {\normalfont\large\bfseries}}
\renewcommand{\subsection}{\@startsection{subsection}{2}{0em}%
        {-3.25ex\@plus -1ex \@minus -.2ex}%
        {1.5ex \@plus .2ex}%
        {\normalfont\bfseries}}
\renewcommand{\subsubsection}%
        {\@startsection{subsubsection}{3}{0em}%
        {-3.25ex\@plus -1ex \@minus -.2ex}%
        {1.5ex \@plus .2ex}%
        {\normalfont\itshape}}
\makeatother

\newcommand{\SNS}{Scuola Normale Superiore and INFN, Sezione di Pisa, \\
I--56126 Pisa, Italy}

\newcommand{\DPNC}{Departement de Physique Nucleaire et 
Corpuscolaire, Universit\'e de Gen\`eve CH--1211 Gen\`eve 4, 
Switzerland}

\newcommand{\titletext}{Effects of new physics in neutrino 
oscillations in matter} 
\newcommand{\authortext}{\large Mario Campanelli$^{\, a}$ and Andrea
Romanino$^{\, b}$
\medskip\\\em\normalsize 
$\mbox{}^a$ \DPNC
\\[0.1\baselineskip] 
$\mbox{}^b$ \SNS}
\newcommand{\abstracttext}{
  A new flavor changing electron neutrino interaction with matter
  would always dominate the $\nu_e$ oscillation probability at
  sufficiently high neutrino energies. Being suppressed by
  $\theta_{13}$, the energy scale at which the new effect starts to be
  relevant may be within the reach of realistic experiments, where the
  peculiar dependence of the signal with energy could give rise to a
  clear signature in the $\nu_e\rightarrow\nu_\tau$ channel.
  The latter could be observed by means of a coarse large magnetized
  detector by exploiting $\tau\rightarrow\mu$ decays. We discuss the
  possibility of identifying or constraining such effects with a high
  energy neutrino factory. We also comment on the model independent
  limits on them.}

\title{
\normalsize
\Large\bfseries\titletext\bigskip}
\author{\begin{minipage}[t]{0.8\textwidth}
\normalsize\centering\authortext
\end{minipage}}
\date{}

\begin{document}

\maketitle
\begin{abstract}\normalsize\noindent
\abstracttext
\end{abstract}\normalsize\vspace{\baselineskip}

\noindent

\section{Introduction}
The Standard Model (SM) provides an elegant minimal framework for
studying neutrino physics. In this framework, neutrinos are produced
and detected, and interact with matter during propagation, through
charged and neutral current renormalizable interactions. At the same
time, the mixing with charged leptons and the (small) masses can be
accounted for by non-renormalizable interactions in the form
$h_{ij}(L_i H)(L_j H)/\Lambda$. Whatever is the ultraviolet completion
of the SM, it would be no surprise if the direct and indirect signals
looked for at colliders and low-energy experiments were accompanied by
effects in neutrino experiments. Indeed, such effects may arise at a
significant level e.g.\ in supersymmetric~\cite{Aulakh:82a} or
extra-dimensional~\cite{Dienes:98a,DeGouvea:01a} new physics
scenarios.

The possibility that new physics (NP) affects the neutrino transitions
observed in
solar~\cite{Wolfenstein:78a,Bergmann:98a,Bergmann:00a,Berezhiani:01b},
atmospheric~\cite{Ma:98a,Fornengo:99a,Bergmann:99a},
LSND~\cite{Bergmann:98c}, and supernova~\cite{Mansour:98a} experiments
has been widely studied in the literature. Here we consider, from a
model independent point of view, the possibility of measurable effects
in a high intensity controlled neutrino experiment.  As in all other
cases, NP can give rise to corrections to the production or detection
interaction~\cite{Bueno:00a,Gonzalez-Garcia:01a,Ota:01a}, as well as
to the interaction with matter during
propagation~\cite{Gago:01a,Huber:01b,Ota:01a,Huber:02a}.  Here we are
interested in the latter possibility. Effects in the production and
detection processes may also play a role, although not necessarily.
However, the features of the two types of effects are quite different,
which should make relatively easy to disentangle them.  In fact, due to
the geometrical $L^{-2}$ suppression, the latter are best studied at a
smaller baseline $L$~\cite{Gonzalez-Garcia:01a}, whereas in the former
case the $L^{-2}$ suppression is compensated (up to a certain $L$) by
the unfolding of the oscillation. Moreover, the former effect exhibits
a peculiar growth with the energy which, as we will see, may give rise
to a noticeable signature.

New (coherent) effects during propagation manifest themselves as an
effective potential contributions to the neutrino squared mass matrix
$\M^2$, the potentially largest one contributing to
$\M^2_{\nu_e\nu_\tau}$~\cite{Bergmann:98a,Bergmann:99a,Bergmann:00a,Berezhiani:01b}.
Due to the large $\nu_\mu$-$\nu_\tau$ mixing ($\theta_{23}$), the
latter will affect both $\nu_e\rightarrow\nu_\mu$ and
$\nu_e\rightarrow\nu_\tau$ oscillations. However, the effect in the
$\nu_e\rightarrow\nu_\tau$ channel is energy enhanced, while the one
in the $\nu_e\rightarrow\nu_\mu$ is not. As a consequence, the
$\nu_\tau$ spectrum may have a striking enhancement at high energy
contrary to the conventional MSW prediction. Direct event-by-event
detection of such an effect (as well as detection of effects in the
$\nu_\mu\rightarrow\nu_\tau$ channel)~\cite{Gago:01a} would require a
very granular detector for $\tau$ identification.  Would a coarse
neutrino detector with only muon charge identification capability miss
the peculiar feature of the signal and possibly confuse it with a pure
oscillation signal (or viceversa)? Not completely, since the
$\nu_e\to\nu_\tau$ channel will contribute to the wrong sign muon
spectrum through $\tau\to\mu$ decay. This has also implications for
the sensitivity of a wrong sign experiment to new physics, which
becomes strongly dependent on the energy.  Such an unequivocal
departure from MSW predictions would represent a clean signal and a
handle to separate the effect from standard oscillations or
corrections to the production or detection interaction.  Another
effect, not considered here since it would require a detector with
ability of distinguishing between electron-like and neutral
current-like events, would be a large increase of the latter sample,
due to hadronic tau decays.

\section{Theoretical background}
\label{sec:th}

The standard MSW effect gives rise to diagonal contributions to the
neutrino squared mass matrix proportional to the neutrino energy. At
energies above the resonance, those terms suppress the electron
neutrino mixing. On the other hand, a flavor changing neutrino
interaction would give rise to a non-diagonal term which, although
smaller than the diagonal ones, will also grow with the neutrino
energy. In the high energy limit, the two matter induced terms will
eventually dominate the mass matrix. Unlike in the conventional MSW
case, where the mixing angle goes to zero, in this case the mixing
angle reaches an asymptotic value which measures the ratio between
the flavor changing and flavor conserving interactions.

The interest of this simple observation depends on the scale at which
the new interaction starts to become relevant. This happens when the
new matter term becomes comparable to the original entry in the
squared mass matrix. Particularly interesting is then a
$\nu_e$-$\nu_\tau$ flavor changing interaction. The original
$\M^2_{\nu_e \nu_\tau}$ term, as well as $\M^2_{\nu_e \nu_\mu}$, is in
fact suppressed by $\theta_{13}$, the only neutrino mixing angle which
is certainly not large. At the same time, the bounds on
$\nu_e$-$\nu_\tau$ interactions are the weakest among all possible
neutrino flavor changing interactions.

Let us discuss the points above in greater detail.  The neutrino
potential induced by $\nu_\alpha f \rightarrow \nu_\beta f$
interactions ($\alpha,\beta=e,\mu,\tau$ and $f$ represents an electron
or a up or down quark) can be parameterized as $V_{\alpha\beta} =
\sqrt{2}G_F \epsilon_{\alpha\beta} N_e = \epsilon_{\alpha\beta}\V$,
where $\V = \sqrt{2}G_F N_e$, $N_e$ is the electron number density and
$\epsilon_{\alpha\beta}=
|\epsilon_{\alpha\beta}|e^{i\phi_{\alpha\beta}}$ are small parameters
satisfying $\epsilon_{\beta\alpha}=\epsilon^*_{\alpha\beta}$ (see the
Appendix for a detailed discussion of the bounds on these parameters).

The effective neutrino squared mass matrix $\Meff^2$ is then
\begin{equation}
  \label{eq:MM}
  \Meff^2 = U 
  \begin{pmatrix}
    0 & 0 & 0 \\
    0 & \dm{21} & 0 \\
    0 & 0 & \dm{31}
  \end{pmatrix} U^\dagger
  + 2 E \V
  \begin{pmatrix}
    1+\epsilon_{ee} &
    \epsilon_{e\mu} &
    \epsilon_{e\tau}  \\
    \epsilon_{\mu e} &
    \epsilon_{\mu\mu} &
    \epsilon_{\mu\tau}  \\
    \epsilon_{\tau e} &
    \epsilon_{\tau\mu} &
    \epsilon_{\tau\tau} 
  \end{pmatrix} \; ,
\end{equation}
where $E$ is the neutrino energy, $U$ is the MNS mixing matrix in the
usual parameterization and $\dm{21}>0$ is the smaller squared mass
difference.

In order to have an intuitive picture of the basic features of the
effects under study, let us first of all set $\dm{21}=0$.  Whereas in
this limit the angle $\theta_{12}$ becomes unphysical, in presence of
new physics the CP-violating phase $\delta$ does not, contrary to what
sometimes stated in the literature. In fact, the phase redefinition
necessary to rotate $\delta$ away from the mixing matrix moves
$\delta$ in the non-diagonal new interactions. In the convention in
which the $\epsilon$ parameters are initially real and the mixing
matrix is complex\footnote{The parameterization choice for the MNS
  matrix fixes the phase convention for the $\epsilon$ parameters as
  well.}, the $\epsilon$'s become complex once the phase has been
rotated away from the mixing matrix.

Having set $\dm{21}=0$ we can rewrite \eq{MM} as
\begin{equation}
  \label{eq:MM2}
  \Meff^2 = \dm{31} 
  \begin{pmatrix}
    |s^2_{13}|^2 + (E/\Eres)(1+\epsilon_{ee}) &
     s_{13}^*/\sqrt{2} + (E/\Eres) \epsilon^*_{\mu e} &
     s_{13}^*/\sqrt{2} + (E/\Eres) \epsilon^*_{\tau e} \\
     s_{13}/\sqrt{2} + (E/\Eres) \epsilon_{\mu e} & 
     1/2 + (E/\Eres) \epsilon_{\mu\mu} & 
     1/2 + (E/\Eres) \epsilon^*_{\tau\mu} \\
     s_{13}/\sqrt{2} + (E/\Eres) \epsilon_{\tau e} & 
     1/2 + (E/\Eres) \epsilon_{\tau\mu} & 
     1/2 + (E/\Eres) \epsilon_{\tau\tau}  
   \end{pmatrix} \; ,
\end{equation}
where we also set $\theta_{23}=\pi/4$,
$\cos\theta_{13},\cos2\theta_{13}=1$ and we denoted $s_{13}\equiv
\sin\theta_{13}e^{i\delta}$. The resonant energy $\Eres$ is
\begin{equation}
  \label{eq:Eres}
  \Eres \simeq 10\GeV \fracwithdelims{(}{)}{\dm{31}}{2.5\cdot
  10^{-3}\eV^2}\fracwithdelims{(}{)}{1.65\,\mathrm{g}\cm^3}{\rho\, Y_e}
  \; ,
\end{equation}
where $Y_e$ is the number of electrons per baryon in matter $n_e/n_B$.
The $E/\Eres$ enhancement of matter effects allows even small
$\epsilon$ parameters to have a role at sufficiently high energy. This
may happen at accessible energies especially in the case of the $\ete$
parameter, which is most weekly constrained --- values of $\ete$ as
large as 0.1 are not excluded, see the Appendix. In fact, for $\ete =
0.1$ and $E=50\GeV$, the NP term in $(\Meff)_{\tau e}$ would
correspond to a maximal $\sin^2 2\theta_{13}$. In general, $\ete$ has
an effect comparable to that of a $\sin\theta_{13}\simeq
7\epsilon\,(E/50\GeV)$. Since machines with a sensitivity to
$\sin\theta_{13}$ as low as $0.5\cdot 10^{-2}$ are conceivable, we
conclude that sensitivities well below $\ete = 0.1$ are in principle
achievable. Of course, the standard oscillation effect could still be
larger and hide the new effect. However, the $\ete$ term, as the
$\etu$ one, has the additional merit of contributing to an entry of
$\Meff$ that is suppressed by $\theta_{13}$ is absence of new physics.
Therefore, the $\ete$ term becomes comparable to and then exceeds the
standard one for energies $E\gtrsim\ENP$, where $\ENP =
|s_{13}/(\sqrt{2}\,\epsilon)|\Eres$ is suppressed by $\theta_{13}$.
The regime where the new effects are comparable to the standard ones
is within the reach of a machine producing neutrinos of maximum energy
$E_{\text{max}}$ if
\begin{equation}
  \label{eq:threshold}
  |\epsilon| \gtrsim \frac{|s_{13}|}{\sqrt{2}}
   \frac{\Eres}{E_{\text{max}}} \; .
\end{equation}
For example, for $E_{\text{max}} = 50\GeV$, that condition becomes
$|\epsilon|\gtrsim 0.007 (|s_{13}|/0.05)$. We recall that
$|s_{13}|=0.05$ corresponds to $\sin^2 2\theta_{13} = 10^{-2}$, a
value one order of magnitude below the present bound and well within a
typical sensitivity.

From the experimental point of view, a study of the effect in the
$\nu_e\leftrightarrow\nu_\tau$ channel through direct $\tau$ detection
is certainly challenging. However, it is possible to take advantage of
$\tau$ decays into muons to look for the effect in the spectrum of
``wrong sign'' muon events by means of a coarse neutrino detector with
only muon identification capabilities (see e.g. ~\cite{Cervera:00a}).
Whereas a sizable $\etu$ might also give rise to some effects, we
focus here on $\ete$ and set all the other $\epsilon_{\alpha\beta}$
parameters to zero in our numerical calculations.

\interskip

Let us now discuss in greater detail the high energy enhancement of
the $\nu_\tau$ spectrum in presence of a sizable $\ete$ and see that
there is not a corresponding enhancement of the $\nu_\mu$ spectrum,
despite the large mixing of $\nu_\mu$ and $\nu_\tau$. To this aim let us
consider the leading terms of the oscillation probabilities in the
high energy limit $E\gg\Eres$ and compare the cases with and without
new physics.

In the standard case (and in the $\dm{21}=0$ limit) $\theta_{12}$ is
unphysical and the $\theta_{23}$ rotation commutes with the matter
term, so matter effects only modify $\theta_{13}$.  Moreover, the
$\nu_e\leftrightarrow\nu_\tau$ ($\nu_e\leftrightarrow\nu_\mu$)
oscillation probability coincides with a two neutrino oscillation
probability in matter with vacuum amplitude
$\cos^2\theta_{23}\sin^22\theta_{13}$
($\sin^2\theta_{23}\sin^22\theta_{13}$) and squared mass difference
$\dm{31}$. In the limit $E\gg\Eres$, the squared mass difference in
matter $\dm{31}'\sim 2E\V$ grows with energy, canceling the $1/E$
dependence in the oscillating term of the probability. However, the
$\theta_{13}$ mixing angle gets suppressed by the large diagonal MSW
term, $\sin^22\theta'_{13}\sim \sin^22\theta_{13} (\Eres/E)^2$, so the
transition probabilities still decrease as $E^2$, \globallabel{eq:he1}
\begin{align}
  P(\nu_e\rightarrow\nu_\tau) & \sim \Big(\frac{\Eres}{E}\Big)^2
  \cos^2\theta_{23}\sin^2 2\theta_{13} \sin^2\frac{L\V}{2}
  \mytag \\
  P(\nu_e\rightarrow\nu_\mu) & \sim \Big(\frac{\Eres}{E}\Big)^2
  \sin^2\theta_{23}\sin^2 2\theta_{13} \sin^2\frac{L\V}{2} \; . \mytag 
\end{align}

In presence of $\nu_e$-$\nu_\tau$ or $\nu_e$-$\nu_\mu$ flavor changing
interactions, the $\theta_{23}$ rotation does not commute with the
matter term anymore. As a consequence, the oscillation probability in
the $\dm{21}=0$ limit has not a simple two neutrino form. Two squared
mass differences enter the oscillation probabilities, $\dm{31}$ and
$2E\V$ in the large $E/\Eres$ limit.  Moreover, whereas the
$\theta_{23}$ is essentially not affected, $s_{13}$ becomes $s'_{13}
\simeq E/\Eres s_{13} +c_{23}\epsilon_{\tau e}+s_{23}\epsilon_{\mu e}$
and a non vanishing (and physical) 12 angle is generated
$\theta'_{12}\simeq -s_{23}\epsilon_{\tau e} +c_{23}\epsilon_{\mu e}$.
Approximate analytical formulae for the oscillation probabilities are
unfortunately quite cumbersome. At the moment, however, we are only
interested to the leading contributions in the large $E/\Eres$ limit,
that have the following simple expression, 
\globallabel{eq:he2}
\begin{align}
\label{eq:he}
  P(\nu_e\rightarrow\nu_\tau) & \sim 4\left| \epsilon_{\tau e} +
    \frac{\Eres}{E} c_{23}s_{13} \right|^2 \sin^2 \frac{L\V}{2} \mytag
    \\
  P(\nu_e\rightarrow\nu_\mu) & \sim4\left| \epsilon_{\mu e} +
    \frac{\Eres}{E} s_{23}s_{13} \right|^2 \sin^2 \frac{L\V}{2} \;,
    , \mytag 
\end{align}
where we included the leading $\Eres/E$ correction to the
energy-independent amplitudes.  We explicitly see that the
oscillation probability is not suppressed any more by two powers of
the energy but reaches instead a constant value
$4|\epsilon|^2\sin^2(L\V/2)$ at high energies. In an experiment at a
Neutrino Factory, the effect of such an energy independent probability
is enhanced by the growth with energy of the neutrino flux and of the
neutrino cross section. This gives rise to a striking (approximately
cubic before approaching the energy cut off) growth of the signal with
energy.

Since $s_{13}=\sin\theta_{13}e^{i\delta}$ and $\epsilon =
|\epsilon|e^{i\phi}$, we also see from \eqs{he2} that in the high
energy limit the amplitude depends on the cosine of the phase
difference $\delta - \phi$. Let us consider a CP-conserving case, so
that both $s_{13}$ and $\epsilon$ can be made real. Then if the two
parameters have the same sign ($\delta-\phi=0$), the contributions to
the amplitude will interfere constructively in the neutrino channel
and destructively in the antineutrino one.  Viceversa, if the signs
are opposite ($\delta-\phi=\pi$), the interference will be destructive
for neutrinos and constructive for antineutrinos. The destructive
interference will be maximal for
$E\sim\ENP=|s_{13}/(\sqrt{2}\epsilon)|\Eres$, the energy at which the
two contributions to the amplitude are comparable.  The latter formula
and previous considerations hold provided that $\ENP>\Eres$, or
$|s_{13}/(\sqrt{2}\epsilon)|>1$, the condition for \eqs{he2} to be
valid at the energy at which the amplitude vanishes.  For
$\ENP\lesssim\Eres$ the cancellation will be spoiled by the $\dm{31}$
terms.

\Eqs{he2} show that $\ete$ only affects $P(\nu_e\rightarrow\nu_\tau)$
at the leading order in $\Eres/E$. Since $\nu_\tau$ and $\nu_\mu$ are
largely mixed, $\epsilon_{\tau e}$ also enters
$\nu_e\leftrightarrow\nu_\mu$ oscillations but only through subleading
terms omitted in eq.~(\ref{eq:he2}b). This is because the mixing takes
place at atmospheric squared mass difference $\dm{31}$, which in the
large $\Eres/E$ high limit is subleading compared to the other squared
mass difference $2E\V$.

\section{Oscillation probabilities}

We can show now some examples of how the oscillation probability gets
modified by the presence of this new interaction. For doing that we
have to consider a specific case. Let us consider then a model with no
CP violation and $s_{13},\ete>0$,\footnote{The parameter $s_{13}$ can
  always been made positive. Then if CP is conserved $\ete$ is real
  but can have both signs.} with oscillation parameters
$\theta_{23}=\pi/4$, $\sin^2 2\theta_{12}=0.08$, $\dm{31}=3\times
10^{-3}$, $\dm{21}=5\times 10^{-5}$ and $\sin^2 2\theta_{13}=0.001$,
about ten times smaller than the present bound. It has to be noticed
that the smaller the value of $\sin^2 2\theta_{13}$, the more visible
new physics effects are.

We consider a Neutrino Factory with $10^{21}$ muon decays, and a 40
kton detector with only muon identification capabilities, located at a
distance of 3000 km from the accelerator.

\begin{figure}
\begin{minipage}[h]{0.47\textwidth}
\begin{center}
\epsfig{file=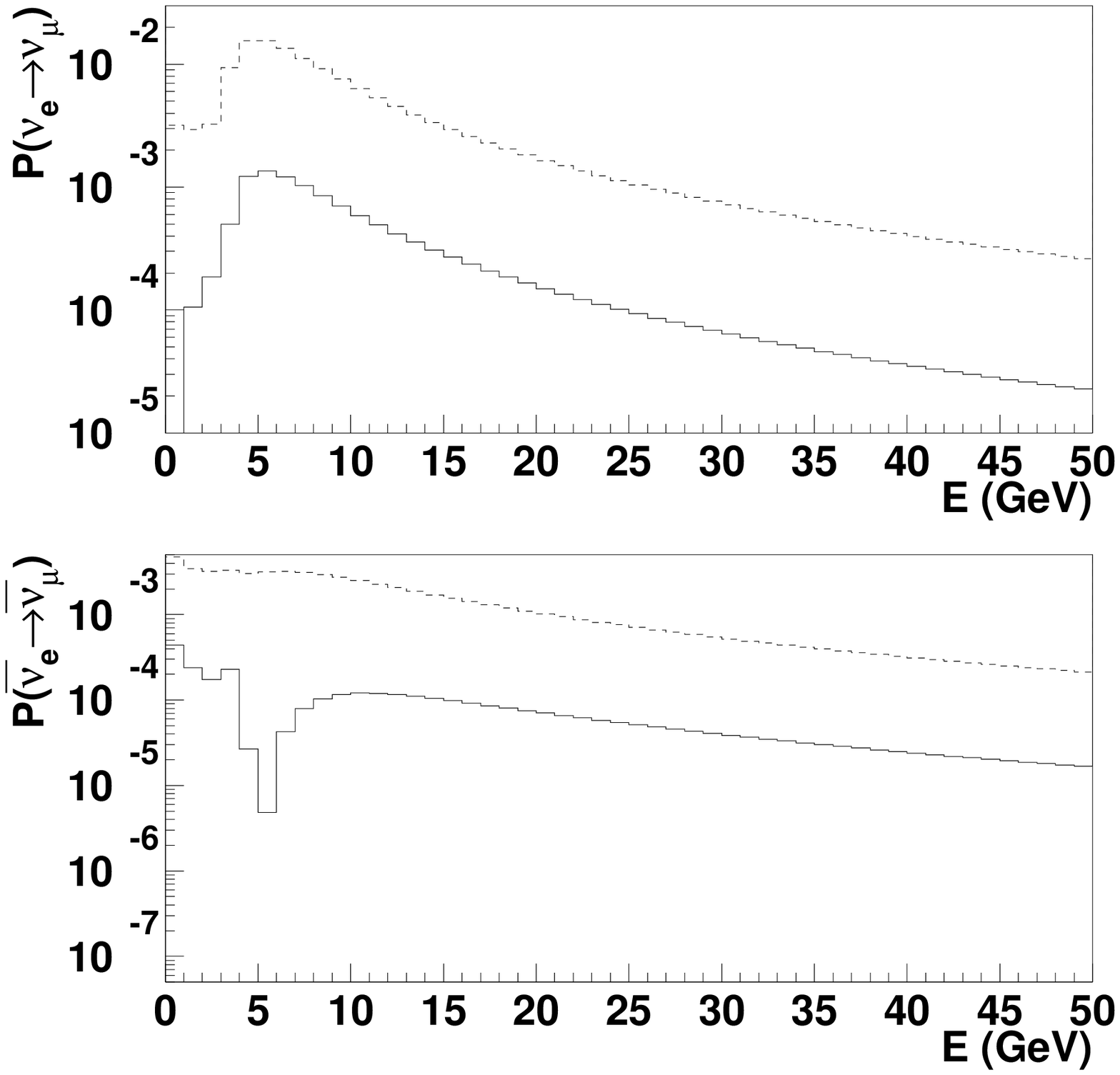,width=8cm}
\end{center}
\caption{$\nu_e\rightarrow\nu_\mu$ oscillation probability in the standard
MSW model (full line) and in the presence of flavor-changing interactions
(dashed line), for $\sin^2 2\theta_{13}=0.001$ and $\ete=0.07$.}
\label{fig:probem}
\end{minipage}
\hfill
\begin{minipage}[h]{0.47\textwidth}
\begin{center}
\epsfig{file=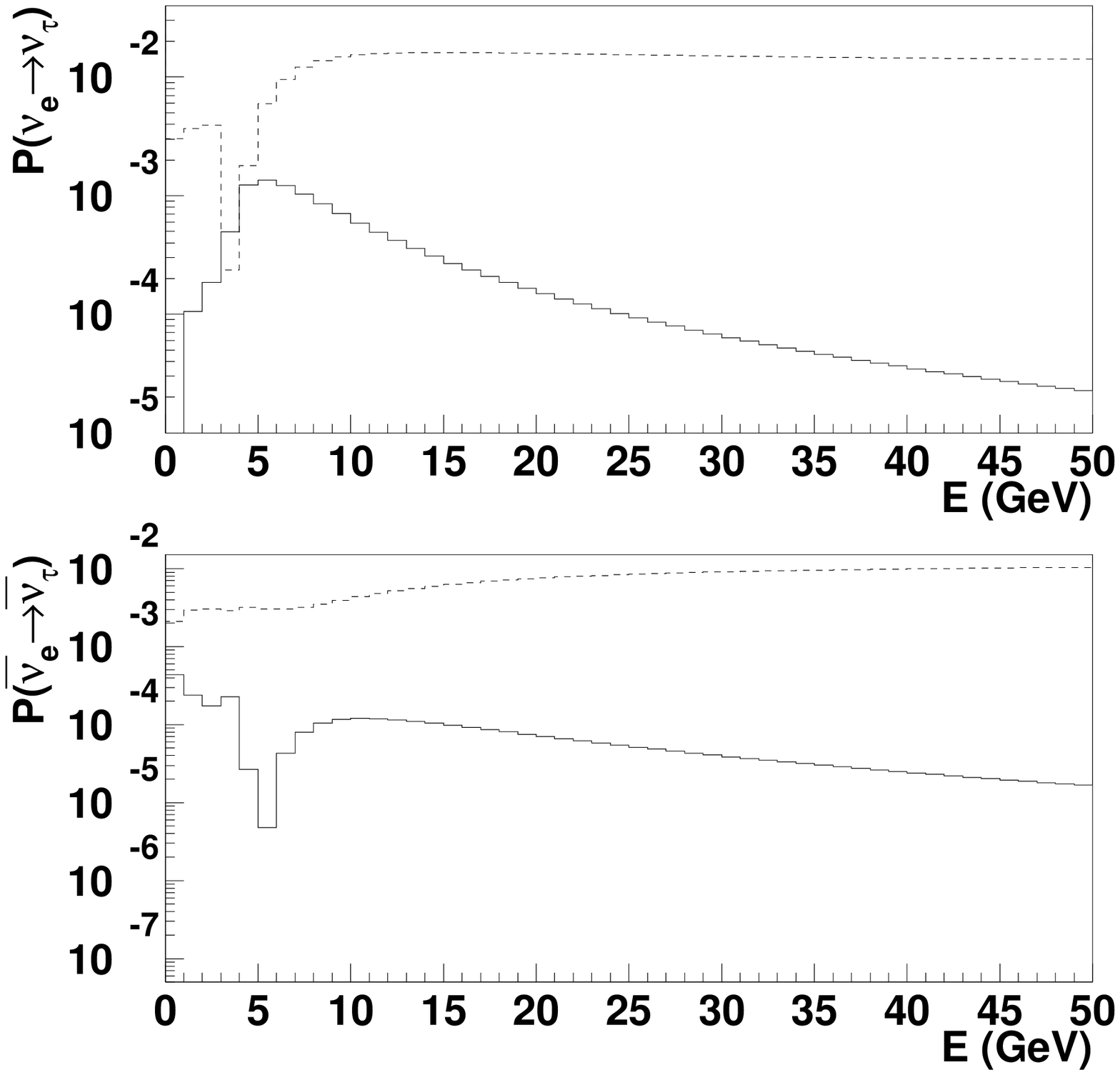,width=8cm}
\end{center}
\caption{$\nu_e\rightarrow\nu_\tau$ oscillation probability in the standard
MSW model (full line) and in the presence of flavor-changing interactions
(dashed line), for $\sin^2 2\theta_{13}=0.001$ and $\ete=0.07$.}
\label{fig:probet}
\end{minipage}
\end{figure}

Present data on Flavor Changing Interactions suggest bounds on the
various elements of the FCI matrix as $|\epsilon_{e\mu}\lesssim
7\times 10^{-5}|$, $|\epsilon_{\mu\tau}\lesssim 5\times 10^{-2}|$,
$|\epsilon_{e\tau}\lesssim 7\times 10^{-2}|$. Even if we assume that
$\eue$ is as large as the experimental bound, its effect on
oscillation probabilities will be negligible. An $\epsilon_{\tau\mu}$
at the experimental bound would give rise to non-negligible
effects~\cite{Gago:01a} but not to the high-energy
enhancement we are focusing on, so we set $\epsilon_{\tau\mu}=0$.
The effect of a sizable $\ete$ is shown in \Fig{probet}, where the
solid histograms represent normal matter oscillations whereas the
dashed histogram is the new physics case. Assuming $\ete$ is close to
the allowed maximum value, the oscillation probability gets modified
in quite a dramatic way.


We see that, while the normal oscillation probability decreases like
$1/E_\nu^2$, in the case of new physics the probability tends to a
constant; the difference is of course enhanced at high energy. For
antineutrinos, we observe the same behavior at high energy, but we
note a difference at intermediate energies $E\sim\ENP$. There, the two
terms in the amplitude of eq.~(\ref{eq:he2}a) are comparable and their
relative sign is opposite for neutrinos and antineutrinos. The effect
of such a sign change is clearly visible but cannot accounted for by
eq.~(\ref{eq:he2}a) for small values of $s_{13}/(\sqrt{2}\ete)$, as in
the present example. On the contrary, for $s_{13}/\ete>\sqrt{2}$ (and
e.g.\ no CP-violation, $s_{13}$, $\ete$ both positive as in this
example) the sign change would lead to a suppression of the
antineutrino probability located at $E \simeq \ENP$, as apparent from
eq.~(\ref{eq:he2}a). In the latter case, the suppression of the
antineutrino oscillation probability measures
$s_{13}/\ete=\sqrt{2}\ENP/\Eres$. In any case, the difference between
the two CP-conjugated channels at $E\sim\ENP$ represents a powerful
tool to constrain the relative phase of $s_{13}$ and $\ete$.

In the $\nu_e\rightarrow\nu_\mu$ channel, the effect of $\ete$ is
quite different, as discussed in Sec.~\ref{sec:th}. At high energy,
where the dominant squared mass difference is $2EV$, the oscillation
probabilities are not enhanced. On the contrary, the spectrum does not
differ from the pure oscillation case apart from the normalization and
falls with energy as in the standard MSW case. The new physics term
behaves as a larger $\sin^2 2\theta_{13}$~\cite{Huber:01b,Huber:02a}.

\section{Observable effects at a Neutrino Factory}

The neutrino energy distribution in muon decays is the following:
\[\frac{d^2 N_{\nu_\mu}}{dx d\Omega}\propto \frac{2
  x^2}{4\pi}[(3-2x)+(1-2x)P_\mu 
\cos\theta]\]
\[\frac{d^2 N_{\bar{\nu}_e}}{dx d\Omega}\propto \frac{12
  x^2}{4\pi}[(1-x)+(1-x)P_\mu \cos\theta]\] 
where $x=2E_\nu/m_\mu$, $P_\mu$ is the muon polarization, and $\theta$
is the angle between the muon polarization vector and the neutrino
direction.  In the laboratory frame, the shape of the energy
distribution is preserved if a Lorentz boost is applied, so if muons
are accelerated to an energy $E_\mu$, the spectral shape will be the
same, with this time $x=E_\nu/E_\mu$. The cleanest experimental
observable to measure neutrino oscillation is the appearance of
wrong-sign muons, i.e. muons observed in the detector with a charge
opposite to those circulating in the storage ring. When positive muons
circulate in the ring, electron neutrinos are produced via
\[\mu^+\to e^+ \bar{\nu}_\mu\nu_e.\]
$\nu_e$ would then oscillate into $\nu_\mu$, seen as negative
(wrong-sign) muons in a far detector. Also $\nu_e\to\nu_\tau$
oscillations will contribute to this channel, via $\tau\to\mu$ decays
(B.R. $\approx 17\%$), that will produce muons of the same sign of
those coming directly from $\nu_e\to\nu_\mu$ oscillations.  We already
saw in the previous section that the presence of new physics in matter
propagation could lead to a noticeable change of the
$\nu_e\to\nu_\tau$ oscillation probability for large neutrino
energies, especially for small values of $\theta_{13}$. This effect
will then be visible in the wrong-sign muon spectrum due to
$\tau\to\mu$ decays. \par
\begin{figure}[tbh]
\begin{minipage}[h]{0.47\textwidth}
\begin{center}
\epsfig{file=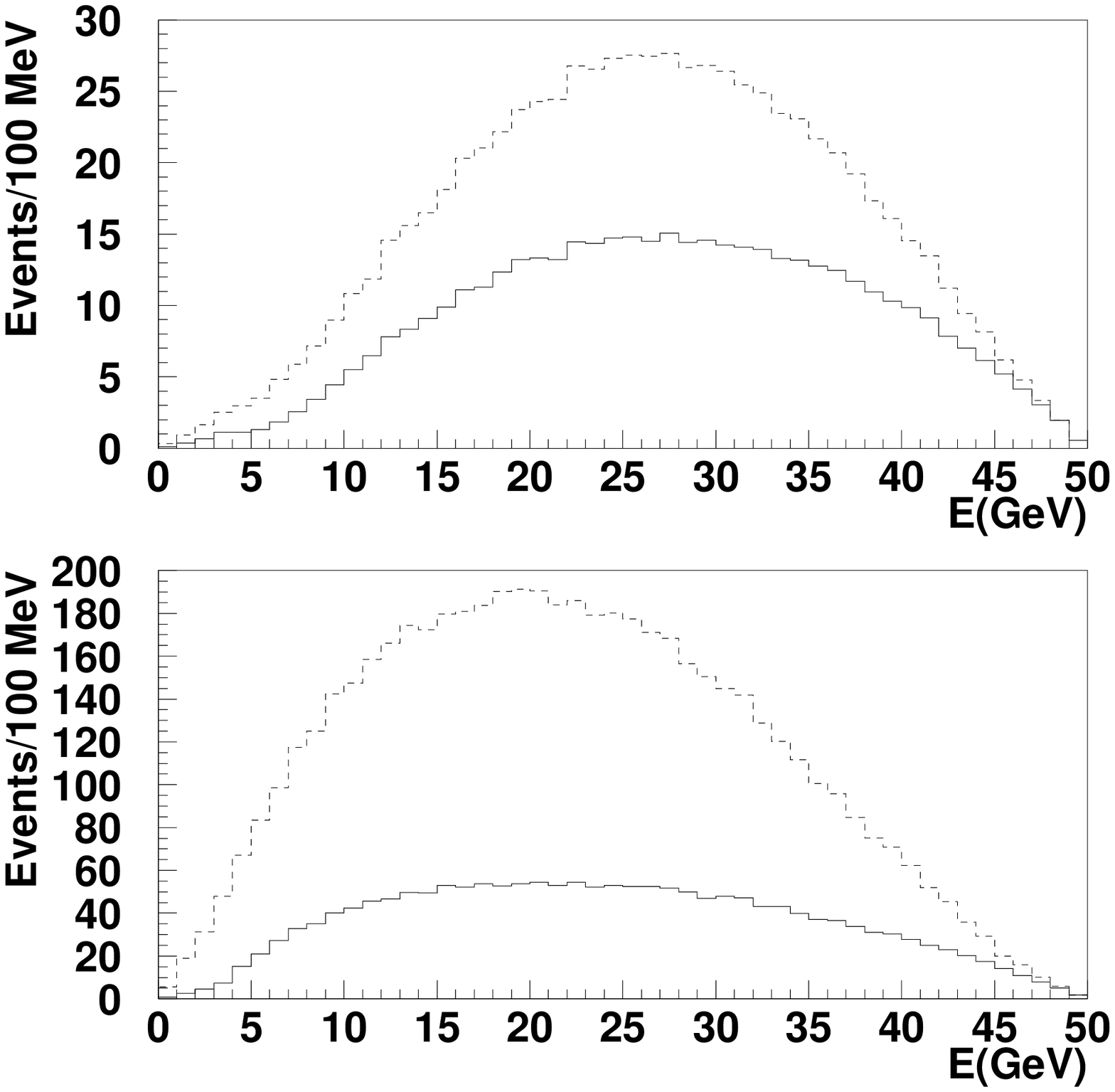,width=8cm}
\end{center}
\caption{Observed visible energy spectrum for wrong-sign muon events
  in the case of $\mu^-$ (upper plot) and $\mu^+$ (lower plot)
  circulating in the storage ring. Full histogram is the standard MSW
  case, dashed histogram is in the presence of new interactions. For
  the considered value of $\sin^2 2\theta_{13}=10^{-2}$, the
  magnitudes of the two effects are similar.}
\label{fig:events-2}
\end{minipage}
\hfill
\begin{minipage}[h]{0.47\textwidth}
\begin{center}
\epsfig{file=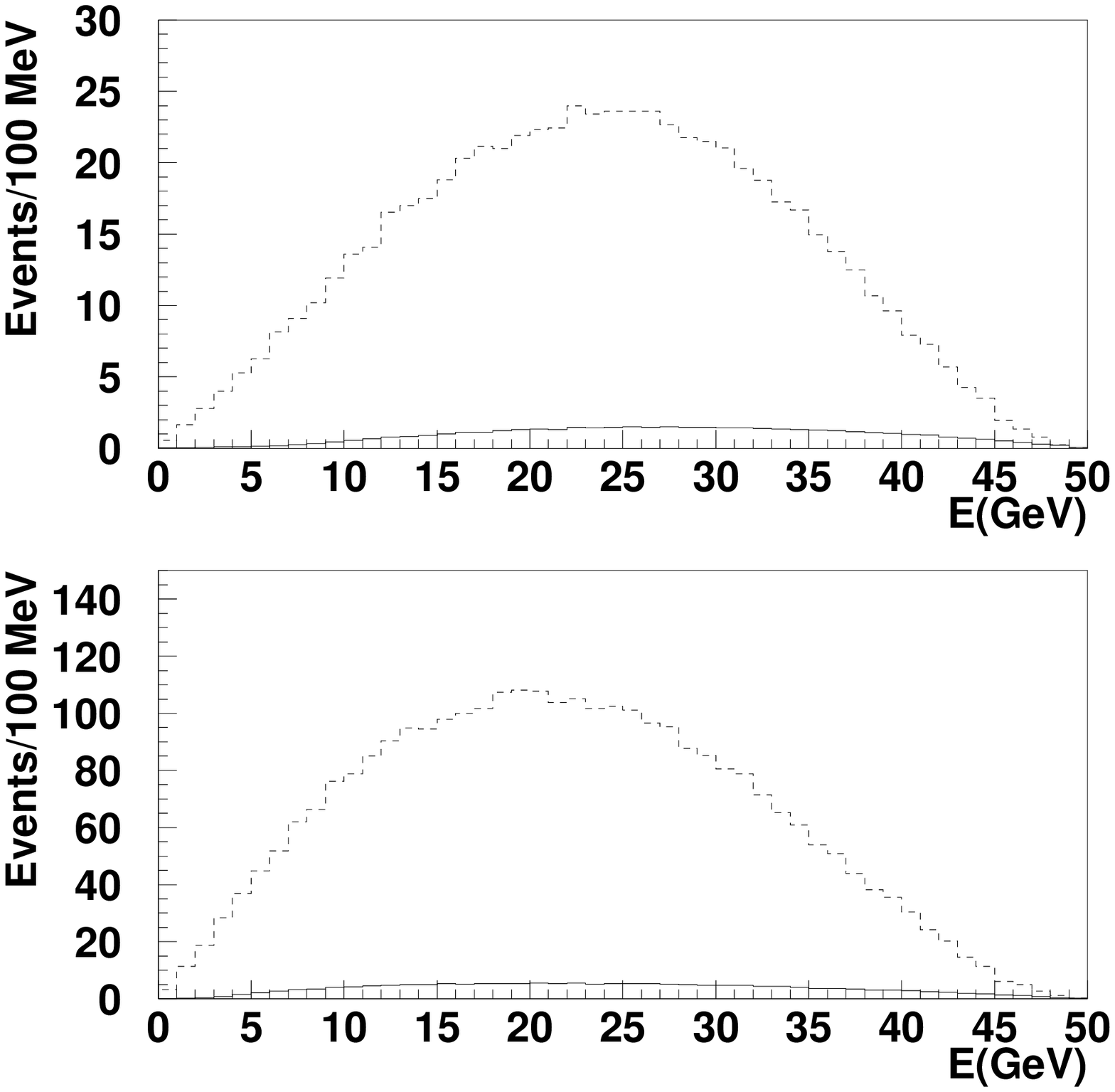,width=8cm}
\end{center}
\caption{Same plot as before, but considering a value of
  $\sin^2 2\theta_{13}=10^{-3}$. As expected, the standard MSW effect
  (full line) is much reduced, therefore the relative importance of
  the new physics interaction (dashed line, almost unchanged with
  respect to the previous case) is largely enhanced.  The two plots
  correspond to $\mu^-$ and $\mu^+$ in the ring.  }
\label{fig:events-3}
\end{minipage}
\end{figure}

 The fact that the neutrino factory flux
increases for large energies allows a precise exploration of the
high-energy oscillation probability. In figures \ref{fig:events-2} and
\ref{fig:events-3} we show spectra for wrong-sign muon events in a
neutrino factory with $10^{21}$ muon decays and a 40 kt far detector.
The large difference observable between normal matter propagation and
FCI at high energy is almost only due to $\tau$ decays, since the
effect of $\ete$ on $\nu_e\to\nu_\mu$ oscillations is is smaller than
in $\nu_e\to\nu_\tau$ and significant only at intermediate energies,
and since the very strong bound existing on $|\epsilon_{e\mu}|$
prevents any measurable effect from this parameter.

In presence of new effects, one wonders whether it is possible to
measure both the new physics parameters, in our case $\ete$, and the
oscillation parameters, in particular $s_{13}$.  The spectral
information is essential for this purpose.  Only counting the number
of wrong-sign muons will lead to the impossibility of correctly
interpreting the observed neutrino transitions in the case of new
physics. The role of $\tau\to\mu$ decays is crucial here. If they were
neglected and only direct $\nu_e\to\nu_\mu$ transitions were
considered, a confusion would arise between $\ete$ and $\theta_{13}$
even if the spectral information were taken into
account~\cite{Huber:01b,Huber:02a}.

\begin{figure}[tbh]
\begin{center}
\epsfig{file=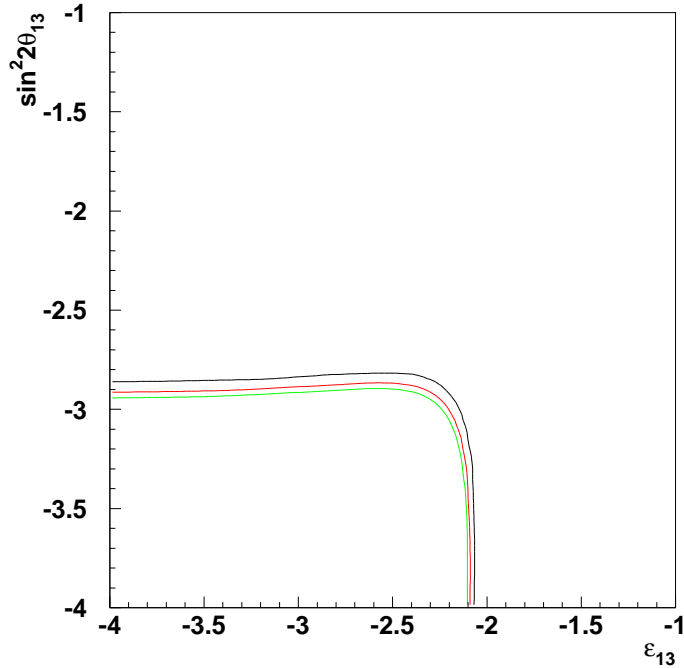,width=0.6\textwidth}
\end{center}
\caption{90\% C.L contour in the plane $\sin^2 2\theta_{13}$,$\epsilon
_{13}$ of the possibility of distinguishing new physics from standard 
oscillations, extracting the oscillation parameter $\theta_{13}$ from
the counting of the number of events and performing a $\chi^2$ test on
the spectral shape.}
\label{fig:contour}
\end{figure}

To be more general, figure \ref{fig:contour} shows 90\% C.L. contours
for observing FCI effects for values of $|\epsilon_{e\mu}|$ in the
range $10^{-4}-10^{-2}$, and $\sin^2 2\theta_{13}$ between $5\times
10^{-5}- 10^{-2}$.\par
To produce this plot we consider an experiment observing a certain 
number of wrong-sign muons in the detector. It is well-known that 
$\theta_{13}$ has quite
a small effect on the energy distribution, therefore can be extracted
from a simple counting of the events. The other relevant parameters will
be measured with very good precision from the ``dip'' of the 
disappearance of right-sign muons (not much affected by new physics 
effects). The experiment makes a 
``measurement'' of $\theta_{13}$ based on event counting, assuming that
no new physics is present, and produces an expected energy
spectrum for that particular value of $\theta_{13}$. Then the spectrum
without new physics is compared
with that actually observed, and a bin-by-bin $\chi^2$ is produced.
If the $\chi^2$ probability of the two histograms to be equal is too 
small, we consider that no confusion can arise between new physics 
effects and normal oscillations.

In obtaining figure \ref{fig:contour} we have set the new CP-violating
phase $\delta-\phi$ to zero. One wonders whether the fact that
$\delta-\phi$ is actually unknown would spoil the results in the
general case. We expect that this is not the case provided that data
with both $\mu^+$ and $\mu^-$ circulating in the ring are used.  In
fact, as mentioned in the previous section, the comparison of the two
sets of data allows in principle to determine the relative sign of
$s_{13}$ and $\ete$ and, more generally, the new CP-violating phase
$\delta-\phi$.\footnote{Notice that the CP-violation effects
  associated to the new physics phase $\delta-\phi$ have a
  characteristic energy scale $E\sim\ENP$ whereas the standard
  CP-violation and other effects related to a $\dm{21}\neq 0$ are
  enhanced at low energy.  This offers an handle to distinguish the
  two sources of CP-violation as well.} Here, we have implicitly
assumed that the information from the comparison of the two sets of
data has been taken into account to constrain $\delta-\phi$ and we
have hence used only one set for discussing the measurement of
$s_{13}$ and $\ete$. A detailed study of CP-violation in this
framework is beyond the scope of this paper.

\section{Conclusions}
We considered here the effects of a possible new flavor-changing
interaction affecting the propagation of electron neutrinos in matter.
Contrary to new effects in neutrino production or interaction, already
well constrained by short-baseline experiments, new physics in matter
propagation would benefit from very long-baselines, like those
considered for a future Neutrino Factory. The characteristic feature
of this machine, i.e.\ the increase of neutrino flux with energy, is
also ideal for these studies since new interactions would not suffer
from the $1/E^2$ suppression, and therefore be much more visible at
high energy.  We studied the problem in a quantitative way, assuming a
coarse magnetized iron detector, and observed that, for new physics
close to the present experimental bounds, a noticeable signal could be
observed in $\nu_e\rightarrow\nu_\tau$ transitions, exploiting the
wrong-sign muons produced in $\tau\rightarrow\mu$ decays.

\section*{Acknowledgments}

M.C. would like to thank the Scuola Normale of Pisa for the kind
ospitality during the last days of writeup of this paper. This work
has been partially supported by MIUR and by the EU under TMR contract
HPRN--CT--2000--00148.

\section*{Appendix}

In this appendix, we review and enlarge the analysis of the limits on
the size of NP contributions to $\Meff$. The general form of new four
fermion operators leading to additional contributions to the neutrino
coherent scattering with an ordinary medium is
\begin{equation}
  \label{eq:L}
  \sum_{
    \substack{f=e,u,d \\ \alpha,\beta = e, \mu, \tau}}
  4\frac{G_F}{\sqrt{2}} 
  \bar\nu_{\alpha} \gamma^\mu \nu_{\beta} \left(
  \epsilon^{f_L}_{\alpha\beta}\,\bar f_L \gamma_\mu f_L +
  \epsilon^{f_R}_{\alpha\beta}\,\bar f_R \gamma_\mu f_R \right) \;.
\end{equation}
Needless to say, \eq{L} is written in a basis in which the kinetic
term is canonical and the charged fermion masses are diagonal. The
$\epsilon$ parameters in \eq{MM} are then given by in terms of the
$\epsilon^f$ in the equation above by
\begin{equation}
  \label{eq:eab}
  \epsilon = \epsilon^e + 2\epsilon^u + \epsilon^d + \frac{n_n}{n_e}
  (2\epsilon^d + \epsilon^u) \;,
\end{equation}
where $n_e$ and $n_n$ are the electron and neutron number densities
respectively and we have omitted the flavor indexes. Notice that the
$\epsilon$ parameters in \eq{MM} have a mild dependence on the distance
traveled through the $n_n/n_e$ ratio.

A model independent limit on $\epsilon_{\tau\mu}$ can be inferred from
atmospheric neutrino data~\cite{Fornengo:99a}: $\epsilon_{\tau
  \mu}\lesssim 0.05$. Significant limits on the single
$\epsilon^{e,u,d}$s can be obtained if one assumes that the operators
in \eq{L} originate from SU(2)$_W$ invariant operators. Then
experimental bounds on charged lepton processes
imply~\cite{Bergmann:98a,Bergmann:99a,Bergmann:00a}
\globallabel{eq:limits}
\begin{align}
  \epsilon^e_{\mu e} & \lesssim 10^{-6} &
  \epsilon^e_{\tau \mu} & \lesssim 3\cdot 10^{-3} &
  \epsilon^e_{\tau e} & \lesssim 4\cdot 10^{-3} \mytag \\
  \epsilon^{u,d}_{\mu e} & \lesssim 10^{-5} &
  \epsilon^{u,d}_{\tau \mu} & \lesssim 10^{-2} &
  \epsilon^{u,d}_{\tau e} & \lesssim 10^{-2} \; . \mytag 
\end{align}
Since SU(2)$_W$ is broken, the limits above can be evaded. To what
extent they can be evaded depends on how the operators \eq{L} are
generated and how SU(2)$_W$ breaking enters. The case of the exchange
of a SU(2)$_W$ multiplet of bosons with SU(2)$_W$ breaking masses has
been considered in~\cite{Bergmann:99a,Bergmann:00a}. In this case, the
operators in \eq{L} are still related to the corresponding charged
lepton operators. However, the limits can be relaxed by the SU(2)$_W$
breaking masses by a factor of about 7 without a conflict with the
electroweak precision data. Alternatively, the effect of SU(2)$_W$
breaking can be studied through an operator expansion in which the
breaking shows up as the vev of the Higgs fields in the higher order
operators~\cite{Berezhiani:01a}.

In both cases above, the limits are inferred from the charged lepton
sector by relating the effects in the two sectors. However, it is
possible to generate the neutrino operator in \eq{L} without giving
rise to any charged lepton effect through a known mechanism not
considered before in this context. In the remainder of this appendix,
we discuss the limits on these possible additional contributions to
\eq{L} from neutrino physics and see that they cannot evade the limits
reviewed above.

Suppose that some physics (e.g.\ warped or flat extra
dimensions~\cite{DeGouvea:01a}) gives rise to the operator
\begin{equation}
  \label{eq:peculiar}
  2\sqrt{2}\ve_{\alpha\beta} G_F (H L_\alpha)^\dagger i\hat\partial(H
  L_\beta) \;.
\end{equation}
Since this operator only contribute to the neutrino wave function,
bringing the neutrino kinetic term back in canonical form will
generate new operators as in \eq{L} but will not generate any
operators involving charged leptons only~\cite{DeGouvea:01a}. The
couplings in \eq{L} generated through this mechanism are
\globallabel{eq:Zren}
\begin{gather}
  \epsilon^e_{\alpha\beta} = -\frac{1}{2} \Bigr(\ve_{\alpha
  e}\delta_{\beta e} + \ve_{e\beta}\delta_{\alpha e}\Bigl) +
  \Bigl(\frac{1}{2} - 2\sin^2\theta_W\Bigr)\ve_{\alpha\beta} \mytag \\
  \epsilon^u_{\alpha\beta} =
  -\Bigl(\frac{1}{2}-\frac{4}{3}\sin^2\theta_W\Bigr)\ve_{\alpha\beta}  \qquad
  \epsilon^d_{\alpha\beta} =
  \Bigl(\frac{1}{2}-\frac{2}{3}\sin^2\theta_W\Bigr)\ve_{\alpha\beta} \;.
\end{gather}
Here the $\ve$ parameters are only constrained by neutrino
experiments, that give~\cite{DeGouvea:01a} $|\ve_{\mu e}| < 0.05$,
$|\ve_{\tau e}| < 0.1$, $|\ve_{\tau \mu}| < 0.013$. In turn, these
bounds imply e.g.  $|\epsilon^e_{\mu e}| < 0.025$, $|\epsilon^e_{\tau
  e}| < 0.05$, to be compared with the much more restrictive
$|\epsilon^e_{\mu e}| < 7\cdot 10^{-6}$, $|\epsilon^e_{\tau e}| <
28\cdot 10^{-3}$ quoted above. Therefore, the operator in \eq{Zren}
gives the potentially largest contribution to $\epsilon^e_{\mu e}$,
$\epsilon^e_{\tau e}$. However, what matters in our case is the value
of $\epsilon_{\alpha\beta}$ rather than the individual
$\epsilon^{e,u,d}_{\alpha\beta}$,
\begin{equation}
  \label{eq:1}
  \epsilon_{\alpha\beta} = -\frac{1}{2}\bigl(\ve_{\alpha
  e}\delta_{\beta e}+\ve_{e\beta}\delta_{\alpha e}\bigr)
  +\frac{1}{2}\frac{n_n}{n_e} \ve_{\alpha\beta} \;.
\end{equation}
In particular, the contribution to the quantity we are interested in,
$\epsilon_{\tau e} = (n_n/n_e-1)\ve_{\tau e}/2$, is suppressed in the
earth by the small $(n_n/n_e-1)/2$ factor. Such a suppression makes
the corresponding bound weaker than those coming from \eqs{limits}.


\begin{thebibliography}{10}

\bibitem{Aulakh:82a}
C.S. Aulakh and R.N. Mohapatra,
\newblock Phys. Lett. B119 (1982) 136. \\
\newblock 
F. Zwirner,
\newblock Phys. Lett. B132 (1983) 103. \\
\newblock 
L.J. Hall and M. Suzuki,
\newblock Nucl. Phys. B231 (1984) 419. \\
\newblock 
J.R. Ellis et~al.,
\newblock Phys. Lett. B150 (1985) 142. \\
\newblock 
G.G. Ross and J.W.F. Valle,
\newblock Phys. Lett. B151 (1985) 375. \\
\newblock 
R. Barbieri and A. Masiero,
\newblock Nucl. Phys. B267 (1986) 679. 
\newblock 

\bibitem{Dienes:98a}
K.R. Dienes, E. Dudas and T. Gherghetta,
\newblock Nucl. Phys. B557 (1999) 25, hep-ph/9811428. \\
\newblock 
N. Arkani-Hamed et~al.,
\newblock (1998), hep-ph/9811448. \\
\newblock 
Y. Grossman and M. Neubert,
\newblock Phys. Lett. B474 (2000) 361, hep-ph/9912408.
\newblock 

\bibitem{DeGouvea:01a}
A. De~Gouvea et~al.,
\newblock Nucl. Phys. B623 (2002) 395, hep-ph/0107156,
\newblock 

\bibitem{Wolfenstein:78a}
L. Wolfenstein,
\newblock Phys. Rev. D17 (1978) 2369. \\
\newblock 
J.W.F. Valle,
\newblock Phys. Lett. B199 (1987) 432. \\
\newblock 
M. Fukugita and T. Yanagida,
\newblock Phys. Lett. B206 (1988) 93. \\
\newblock 
E. Roulet,
\newblock Phys. Rev. D44 (1991) 935. \\
\newblock 
M.M. Guzzo, A. Masiero and S.T. Petcov,
\newblock Phys. Lett. B260 (1991) 154. \\
\newblock 
V.D. Barger, R.J.N. Phillips and K. Whisnant,
\newblock Phys. Rev. D44 (1991) 1629. \\
\newblock 
S. Degl'Innocenti and B. Ricci,
\newblock Mod. Phys. Lett. A8 (1993) 471. \\
\newblock 
G.L. Fogli and E. Lisi,
\newblock Astropart. Phys. 2 (1994) 91. \\
\newblock 
P.I. Krastev and J.N. Bahcall,
\newblock {S}anta {M}onica 1997, {F}lavor-changing neutral currents, pp.
  259--263, 1997, hep-ph/9703267.
\newblock 

\bibitem{Bergmann:98a}
S. Bergmann,
\newblock Nucl. Phys. B515 (1998) 363, hep-ph/9707398.
\newblock 

\bibitem{Bergmann:00a}
S. Bergmann et~al.,
\newblock Phys. Rev. D62 (2000) 073001, hep-ph/0004049.
\newblock 

\bibitem{Berezhiani:01b}
Z. Berezhiani, R.S. Raghavan and A. Rossi,
\newblock (2001), hep-ph/0111138.
\newblock 

\bibitem{Ma:98a}
E. Ma and P. Roy,
\newblock Phys. Rev. Lett. 80 (1998) 4637, hep-ph/9706309. \\
\newblock 
G. Brooijmans,
\newblock (1998), hep-ph/9808498. \\
\newblock 
M.C. Gonzalez-Garcia et~al.,
\newblock Phys. Rev. Lett. 82 (1999) 3202, hep-ph/9809531. \\
\newblock 
P. Lipari and M. Lusignoli,
\newblock Phys. Rev. D60 (1999) 013003, hep-ph/9901350. \\
\newblock 
N. Fornengo et~al.,
\newblock Phys. Rev. D65 (2002) 013010, hep-ph/0108043.
\newblock 

\bibitem{Fornengo:99a}
N. Fornengo, M.C. Gonzalez-Garcia and J.W.F. Valle,
\newblock jhep 07 (2000) 006, hep-ph/9906539.
\newblock 

\bibitem{Bergmann:99a}
S. Bergmann, Y. Grossman and D.M. Pierce,
\newblock Phys. Rev. D61 (2000) 053005, hep-ph/9909390.
\newblock 

\bibitem{Bergmann:98c}
S. Bergmann and Y. Grossman,
\newblock Phys. Rev. D59 (1999) 093005, hep-ph/9809524.
\newblock 

\bibitem{Mansour:98a}
S. Mansour and T.K. Kuo,
\newblock Phys. Rev. D58 (1998) 013012, hep-ph/9711424. \\
\newblock 
S. Bergmann and A. Kagan,
\newblock Nucl. Phys. B538 (1999) 368, hep-ph/9803305. \\
\newblock 
G.L. Fogli et~al.,
\newblock (2002), hep-ph/0202269.
\newblock 

\bibitem{Bueno:00a}
A. Bueno et~al.,
\newblock JHEP 06 (2001) 032, hep-ph/0010308. \\
\newblock 
I.I.Y. Bigi et~al.,
\newblock (2001), hep-ph/0106177. \\
\newblock 
NuTeV, J.A. Formaggio et~al.,
\newblock Phys. Rev. Lett. 87 (2001) 071803, hep-ex/0104029. \\
\newblock 
A. Datta et~al.,
\newblock Phys. Rev. D64 (2001) 015011, hep-ph/0011375.
\newblock 

\bibitem{Gonzalez-Garcia:01a}
M.C. Gonzalez-Garcia et~al.,
\newblock Phys. Rev. D64 (2001) 096006, hep-ph/0105159.
\newblock 

\bibitem{Ota:01a}
T. Ota, J. Sato and N.a. Yamashita,
\newblock (2001), hep-ph/0112329.
\newblock 

\bibitem{Gago:01a}
A.M. Gago et~al.,
\newblock Phys. Rev. D64 (2001) 073003, hep-ph/0105196. \\
\newblock 
P. Huber and J.W.F. Valle,
\newblock Phys. Lett. B523 (2001) 151, hep-ph/0108193. 
\newblock 

\bibitem{Huber:01b}
P. Huber, T. Schwetz and J.W.F. Valle,
\newblock (2001), hep-ph/0111224.
\newblock 

\bibitem{Huber:02a}
P. Huber, T. Schwetz and J.W.F. Valle,
\newblock (2002), hep-ph/0202048.
\newblock 

\bibitem{Cervera:00a}
A. Cervera et~al.,
\newblock Nucl. Phys. B579 (2000) 17, hep-ph/0002108.
\newblock 

\bibitem{Berezhiani:01a}
Z. Berezhiani and A. Rossi,
\newblock (2001), hep-ph/0111137.
\newblock 

\end{thebibliography}

\end{document}